\documentclass{elsart}
\def\lsim{\mathrel{\rlap{\lower4pt\hbox{\hskip1pt$\sim$}}
    \raise1pt\hbox{$<$}}}         
\def\gsim{\mathrel{\rlap{\lower4pt\hbox{\hskip1pt$\sim$}}
    \raise1pt\hbox{$>$}}}         
 \usepackage{graphicx}

\usepackage{amssymb}

\begin{document}
\begin{frontmatter}
\title{Solar Hydrogen Burning and Neutrinos}

\author{W.C.~Haxton$^{1}$, P.D.~Parker$^{2}$, C.E.~Rolfs$^{3}$}

\address{$^{1}$ INT and  Physics Department, Univ. Washington, Seattle, WA,
98195--1550 USA\\
$^{2}$ Physics Department, Yale University, New Haven, CT,
06520--8124 USA\\
$^{3}$ Experimentalphysik III, Ruhr-Universit\"{a}t Bochum,
D--44780 Bochum, Germany}

\begin{abstract}
We summarize the current status of laboratory measurements of nuclear cross
sections of the pp chain and CN cycle.  We discuss the connections between
such measurements, predictions of solar neutrino fluxes, and the conclusion
that solar neutrinos oscillate before reaching earth.
\end{abstract}

\end{frontmatter}

\section{Introduction}
The probable role of thermonuclear reactions in generating energy
within the solar interior was recognized as early as the 1920s by
astrophysicists such as Eddington \cite{1}. This deduction came
about by default, rather than by any direct laboratory observation
of the reactions. Once the earth's age was established to be at
least a few billion years old (see Burchfield \cite{2} for an
interesting history of the development of this idea), it became
apparent that gravitational and chemical energy sources could not
sustain solar evolution, given the known mass, radius, and
luminosity of the sun. In contrast, newly discovered nuclear
reactions and decays could readily supply the required energy over
the requisite time. A key step was taken by Gamow \cite{3}, who
recognized that tunnelling under the Coulomb barrier could enable
nuclear reactions to occur at the low energies typical of the
solar interior, where \textit{kT} $\sim$ a few keV (\textit{e.g.},
Atkinson $\&$ Houtermans \cite{4}). This idea and subsequent
investigations led to descriptions of possible reaction chains for
stellar energy generation. By the end of the 1930s Weizsacker
\cite{5,6}, Bethe $\&$ Critchfield \cite{7}, and Bethe \cite{8}
had described the essential features of the carbon-nitrogen cycle
and the proton--proton chain, the processes by which the sun and
similar stars convert hydrogen to helium.

The neutrino has a similar vintage, and was connected early on to
stellar reactions. In December 1930 Wolfgang Pauli proposed that
the emission of an unobserved spin--1/2 neutral particle -- the electron
neutrino ($\nu_e$) -- might explain the apparent lack of energy
conservation in nuclear beta decay

\[ (A, Z) \to  (A, Z-1) + e^+ + \nu_e \]

Enrico Fermi was present at a number of Pauli's presentations and
discussed the neutrino with him on these occasions. In 1934,
following closely Chadwick's discovery of the neutron, Fermi
proposed a theory of beta decay based on Dirac's description of
electromagnetic interactions, but with weak currents interacting
at a point, rather than at long distance through the
electromagnetic field. He described beta decay as a proton
decaying to a neutron, a phenomenon energetically possible because
of nuclear binding energies, with the emission of a positron and
neutrino. Apart from the absence of parity violation, which would
not be discovered until 1957, Fermi's description is a correct
low-energy approximation to our current standard model of weak
interactions.

Thus, as nuclear astrophysicists unravelled the stellar processes
for hydrogen burning, stars were recognized to be copious sources
of neutrinos

\[4p \to {}^4\rm{He} + 2e^+ + 2\nu_e\]

This observation led Ray Davis to construct the chlorine
experiment to measure the flux of solar neutrinos. The resulting
solar neutrino puzzle was ultimately traced to new physics
(neutrino oscillations) requiring the mixing of massive neutrinos,
a phenomenon beyond the standard electroweak model. This discovery
depended critically on the reliability of solar model neutrino
flux predictions, and thus on the quality of our understanding of
the nuclear reaction rates governing the pp and CN cycles.

The goal of this paper is to summarize our understanding of solar
hydrogen burning, with special emphasis on the connections between
laboratory astrophysics---measurement of nuclear cross sections at
or near energies characteristic of the solar core---and the
predictions of solar neutrino fluxes.

\section{Hydrogen Burning}
Observations of stars reveal a wide variety of stellar conditions,
with luminosities relative to solar spanning a range $L \sim$
10$^{-4}$ to 10$^6$ $L_\odot$ and surface temperatures $T_s
\sim$2000--50000 K. The simplest relation one could propose
between luminosity $L$ and $T_{s}$ is that for a blackbody

\begin{equation}
    L \, = \, 4 \pi R^2 \sigma \,  T_s^4 \Rightarrow L/L_{\odot} \, = \,
    (R/R_\odot)^2 \, (T_s/T_\odot)^4
        \label{eq1}
\end{equation}

\noindent which suggests that stars of a similar structure might
lie along a one--parameter path (corresponding to $R/R_\odot$
above) in the luminosity (or magnitude) vs.~temperature (or color)
plane. In fact, there is a dominant path in the
Hertzsprung--Russell color--magnitude diagram along which roughly
80$\%$ of the stars reside. This is the main sequence, those stars
supporting themselves by hydrogen burning through the pp chain or
CN cycles.

As one such star, the sun is an important test of our theory of
main sequence stellar evolution:~its properties~--~age, mass,
surface composition, luminosity, and helioseismology~--~are by far
the most accurately known among the stars. The standard solar
model (SSM) traces the evolution of the sun over the past 4.6
billion years of main sequence burning, thereby predicting the
present--day temperature and composition profiles, the relative
strengths of competing nuclear reaction chains, and the neutrino
fluxes resulting from those chains. Standard solar models share
four basic assumptions:

\begin{itemize}

\item The sun evolves in hydrostatic equilibrium, maintaining a
local balance between the gravitational force and the pressure
gradient. To describe this condition in detail, one must specify
the equation of state as a function of temperature, density, and
composition.

\item Energy is transported by radiation and convection. While
the solar envelope is convective, radiative transport dominates
in the core region where thermonuclear reactions take place. The
opacity depends sensitively on the solar composition, particularly
the abundances of heavier elements.

\item Thermonuclear reaction chains generate solar energy. The
standard model predicts that over 98$\%$ of this energy is
produced from the pp chain conversion of four protons into $^4$He
(see Fig.~1), with proton burning through the CN cycle
contributing the remaining 2$\%$ (Fig.~2). The Sun is a large but
slow reactor:~the core temperature, $T_c \sim $1.5 $\times$ 10$^7$
K, results in typical center-of-mass energies for reacting
particles of $\sim$ 10 keV, much less than the Coulomb barriers
inhibiting charged particle nuclear reactions. Thus reaction cross
sections are small: in most cases, laboratory measurements are
only possible at higher energies, so that cross section data must
be extrapolated to the solar energies of interest.

\item The model is constrained to produce today's solar radius,
mass, and luminosity. An important assumption of the standard
model is that the Sun was highly convective, and therefore uniform
in composition, when it first entered the main sequence. It is
furthermore assumed that the surface abundances of metals (nuclei
with A $>$ 5) were undisturbed by the subsequent evolution, and
thus provide a record of the initial solar metalicity. The
remaining parameter is the $^4$He/H ratio, which is
adjusted until the model reproduces the present solar luminosity
after 4.6 billion years of evolution. The resulting $^4$He/H mass
fraction ratio is typically 0.27 $\pm$ 0.01, which can be compared
to the big--bang value of 0.23 $\pm$ 0.01. Note that the Sun was
formed from previously processed material.

\end{itemize}

The model that emerges is an evolving Sun. As the core's chemical
composition changes, the opacity and core temperature rise,
producing a 44$\%$ luminosity increase since its arrival on the
main sequence. The temperature rise governs the competition
between the three cycles of the pp chain:~the ppI cycle dominates
below about 1.6 $\times$ 10$^7$ K;~the ppII cycle between
(1.7--2.3) $\times$ 10$^7$ K;~and the ppIII above 2.4 $\times$
10$^7$ K. The central core temperature of today is about 1.55
$\times$ 10$^7$ K.

This competition between the cycles determines the pattern of
neutrino fluxes. One consequence of the thermal evolution of our
sun is the relatively recent appearance of a significant $^8$B
neutrino flux, the species that dominated the Davis experiment. As
this flux is produced in the ppIII cycle, it is a very sensitive
function of the solar core. The SSM predicts an exponential
increase in this flux with a doubling period of about 0.9 billion
years.

A final aspect of SSM evolution is the formation of composition
gradients on nuclear burning timescales. Clearly there is a
gradual enrichment of the solar core in $^4$He, the ashes of the
pp--chain. Another element, $^3$He, is produced and then consumed
in the pp chain, eventually reaching some equilibrium abundance.
The time--scale for equilibrium to be established as well as the
eventual equilibrium abundance are both sharply decreasing
functions of temperature, and thus increasing functions of the
distance from the center of the core. Thus a steep $^3$He density
gradient is established over time.

The SSM has had some notable successes. From helioseismology the
sound speed profile $c(r)$ has been very accurately determined for
the outer 90$\%$ of the Sun, and is in good agreement with the
SSM. (However, this conclusion depends in part on inputs to the
SSM:~recent calculations employing one set of revised opacities do
show some significant discrepancies, as we will discuss later.)
Helioseismology probes important predictions of the SSM, such as
the depth of the convective zone. The SSM is not a complete model
in that it does not attempt to describe from first principles
aspects such as convection that extend beyond 1D descriptions. For
this reason certain features of solar structure, such as the
depletion of surface Li by two orders of magnitude, are not
reproduced. Li destruction is often attributed to convective
processes that operated at some epoch in our sun's history,
dredging Li to a depth where burning takes place.

\section{Reaction Rate Formalism}
\subsection{Nuclear rates and S-factors}
Perhaps the key step that allowed quantitative modelling of the
sun and other main--sequence stars was the description of the
relevant nuclear reaction chains. This required both the
development of a theory for sub--barrier fusion reactions and
careful laboratory measurements to constraint the rates of these
reactions at stellar temperatures.

At the temperatures and densities in the solar interior
(\textit{e.g.}, $T_c \sim$ 15 $\times$ 10$^6$ K and $\rho_c
\sim$ 153 g/cm$^3$ \cite{9}), interacting nuclei reach a
Maxwellian equilibrium distribution in a time that is
infinitesimal compared to nuclear reaction time scales. Therefore,
the reaction rate between two nuclei can written as
(\textit{e.g.}, Burbidge \textit{et al.}, \cite{10};~Clayton
\cite{11}; Rolfs $\&$ Rodney \cite{12}):

\begin{equation} \label{eq2}
r_{12} \, = \, \frac{n_1 \, n_2}{1 + \delta_{12}} \, \langle
{\sigma v} \rangle_{12}.
\end{equation}

Here the Kronecker delta prevents double counting
in the case of identical particles, 
$n_1$ and $n_2$ are the number densities of nuclei of type 1
and type 2 (with atomic number $Z_1$ and $Z_2$, and atomic mass
$A_1$ and $A_2)$, and $\left\langle {\sigma v} \right\rangle
_{12}$ denotes the product of the reaction cross--section and the
relative velocity $v$ of the interacting nuclei, averaged over the
collisions in the stellar gas,

\begin{equation}
\langle \sigma v  \rangle_{12} \, = \, \int_0^\infty \, [ \, \sigma \,
(v) \, \cdot \, v ] \, \Phi \, (v)\, \mbox{d} v.
    \label{eq3}
\end{equation}

Under solar conditions nuclear velocities are
very well approximated by a Maxwell--Boltzmann distribution. It
follows that the relative velocity distribution is also a
Maxwell--Boltzmann, governed by the reduced mass $\mu$ of the
colliding nuclei,

\begin{equation}
\Phi \, (v) \mbox{ d} v = (\frac{\mu}{2 \pi k T})^{3/2} \mbox{
exp} (-\frac{\mu v^2}{2kT}) 4 \pi v^2 \mbox{ d} v.
       \label{eq4}
\end{equation}

Therefore,

\begin{equation}
\langle{\sigma  v}\rangle_{12} \, = \, [ \, \frac{8}{\pi \mu (kT)^3} \,
]^{1/2} \, \int_0^\infty \, E \, \sigma \, (E) \mbox{ exp} \, (\,
-\frac{E}{kT} \, ) \mbox{ d} E,
        \label{eq5}
\end{equation}

where $E$ is the relative kinetic energy in the center--of--mass
system. In order to evaluate $\left\langle {\sigma v}
\right\rangle$ the energy dependence of the reaction cross-section
must be determined. (In Fowler \textit{et al}. \cite{13} and more
recently in Angulo \textit{et al.} \cite{14}, the appropriate
expressions for $\left\langle {\sigma v} \right\rangle$ are
tabulated for a variety of specific reactions, including both
non-resonant and resonant neutron-induced and
charged-particle-induced reactions.)

Almost all of the nuclear reactions relevant to solar energy generation
are nonresonant and charged--particle induced.
For such reactions, as the energy dependence at low energies is
dominated by the Coulomb barrier, cross sections are typically
expressed in terms of the S--factor,

\begin{equation}
\label{eq6} \sigma \left( E \right)\;=\;\frac{S\left( E
\right)}{E}\;\;\mathrm{exp}\left( {-2\pi \eta } \right),
\end{equation}

\noindent where the 1/$E$ term comes from the $\lambda ^{2}$
geometrical factor, exp$\left( {-2\pi \eta }
\right)$ is the usual Gamow barrier penetration term with
$\eta$ = $Z_1 Z_2$ $\alpha$/$v$,
$v$ is the relative velocity, and $\alpha \sim$ 1/137 is the fine structure constant.
(We take $\hbar$ = c = 1.)
That is, the sharp energy dependence associated with s--wave
interactions of point nuclei has been removed from $\sigma (E)$,
leaving a quantity $S(E)$ that contains the nontrivial nuclear
physics, yet is simpler to model over an energy range spanning
both solar reactions and terrestrial laboratory measurements. For
nonresonant reactions $S(E)$ will be a slowly varying function of
$E$.

One must take into account differences in the atomic environments
to correctly relate laboratory measurements of
\textit{$\sigma$}$(E)$ and $S(E)$ to the corresponding quantities in
the solar interior. As light nuclei in the solar core are almost
completely ionized, the solar electron screening correction $f_0$
can be treated in a weak--screening approximation \cite{15}.
The correction depends on the ratio of the Coulomb potential
at the Debye radius $R_D$ to the temperature,

\begin{equation}
    f_0 =\;\mathrm{exp} \left({Z_1 \,Z_2 \;\alpha \over R_D k T}\right)
    =\;\mathrm{exp} \left( {0.188\,Z_1 \,Z_2 \;\zeta \;\rho_0
    ^{1/2}\;T_6^{-3/2} } \right),
        \label{eq7}
\end{equation}

where $\zeta R_D = \left( {k T \over 4 \pi \alpha \rho} \right)^{1/2}$,
$\rho$ is the number density of nucleons, $\rho_0$ is a dimensionless density
measured in g/cm$^3$,
$\zeta$ = [$\sum\limits_i$ ($X_i \,
\frac{Z_i^2}{A_i}$ +  $X_i$; $\frac{Z_i}{A_i}$) ] $^{1/2}$,
$X_i$ is the mass fraction of nuclei of type $i$,
and $T_6$ is the dimensionless temperature in units of 10$^6$ K.
Note that the screening factor enhances solar cross sections.

Substituting (\ref{eq4}) into (\ref{eq1}) yields

\begin{equation}
\label{eq8} \left\langle {\sigma v } \right\rangle_{12} =\;\left[
{\frac{8}{\pi \mu (kT)^3}} \right]^{1/2}f_0 \;\int\limits_0^\infty
{\;S\left( E \right)} \;\exp \left( {-2\pi \eta } \right)\;\exp
\left( {-E/kT} \right)\;\mathrm{d} E.
\end{equation}

The energy dependence of this integrand is shown schematically in
Fig. 3. In order to evaluate this integral we expand $S(E)$ in a
Taylor series,

\begin{equation}
\label{eq9} S(E)=S(0) + ES^{ \prime } (0) + \frac{1}{2}E^{2 }S''
(0) + .......
\end{equation}

Substituting this into (\ref{eq6}) yields (Bahcall \cite{16}),
\begin{eqnarray}
\left\langle {\sigma v} \right\rangle_{12} =\;\left( {\frac{2}{\mu
 kT} }\right)^{1/2}\left( {\Delta \,E_0/kT }
\right)\left( {f_0 \,S_{eff}
} \right)\;\exp \left( {-3E_0 /kT} \right) \hspace*{1.5 cm}  \\
 =1.301 \times 10^{-14} \mathrm{cm}^3/\mathrm{s}\left( {\frac{Z_1 \,Z_2
}{A}} \right)^{1/3}f_0 \;{S_{eff} \over MeV-b} \;T_9^{-2/3} \exp
\left( {-3E_0 /kT} \right) \nonumber
    \label{eq10}
\end{eqnarray}
where
\[ E_0/kT = (\pi Z_1 Z_2 \alpha/ \sqrt{2})^{2/3} (\mu c^2/kT)^{1/3}, \]
\[\Delta E_0/kT = 4[ E_0/ 3 kT]^{1/2}, ~~~~~~~
A={A_1 A_2 \over A_1+A_2}, \]
and
\[S_{eff} = S(0) [1 + \frac{5kT}{36E_0} ] +
S^{\prime} (0) E_0 [1 + \frac{35kT}{36E_0} ] + \frac{1}{2}
S^{''} (0) E_0^2 [1 + \frac{89kT}{36E_0}]. \]

\noindent $E_0$ corresponds to the maximum of the integrand, the
Gamow peak, and is thus the most probable energy of reacting
nuclei (see Fig.~3). $\Delta E_{0}$ corresponds to the full width
of the integrand at 1/$e$ of its maximum value. (For example, for
a typical solar core temperature of $T_{6}$ = 16, $E_{0}$ = 23.4
keV and $\Delta E_{0 }$= 13.1 keV for the $^{3}$He($\alpha
$,$\gamma)^{7}$Be reaction in the proton-proton chain.)

Numerous laboratory measurements have been carried out to
determine $S$(0), $S'$(0), and $S''$(0) for the various nuclear
reactions contributing to the proton--proton chain and the CN
cycle. At the present time it is not technically feasible to
measure the cross sections for most of these reactions in the
region of $E_{0}$, as the cross sections are severely suppressed
by the Coulomb barrier. (Clearly, as the basic timescale is the
solar age, the sun is a very slow reactor operating at a
temperature where many reactions are highly suppressed.) Therefore
$S_{eff}$ often must be determined from extrapolations from
laboratory measurements at higher energies, typically $E \gsim$ 100
to 200 keV.

So far, the one significant reaction for which it has been
possible to measure the cross section down to $E_0$ is the
$^{3}$He($^{3}$He,2p)$^{4}$He reaction~--~see Section 4.3, below.
At the sun's center, $E_{0} \sim$ 22 keV for this reaction. The
measured cross section is $\sigma (E_0) \sim $1.5 pb. This can
be compared with the cross sections for the (p,$\gamma)$ and
($\alpha $,$\gamma )$ capture reactions that contribute in the pp
chain or CN cycle, \textit{e.g.},

\hspace*{1 cm} $^{3}$He($\alpha $,$\gamma )^{7}$Be $\rightarrow
E_{0}$ $\sim$ 23 keV $\rightarrow \sigma (E_0)$ $\sim $ 3 $\times $
10$^{-5}$ pb

\hspace*{1 cm} $^{7}$Be(p,$\gamma)^8$B $\rightarrow E_0$ $\sim$ 18.4
keV $\rightarrow \sigma$($E_0) \sim $ 1.5 $\times $ 10$^{-3}$ pb

\hspace*{1 cm} $^{14}$N(p,$\gamma )^{15}$O $\rightarrow E_0$ $\sim$
27.2 keV $\rightarrow \sigma (E_0) \sim $ 2.2 $\times $ 10$^{-7}$
pb

Direct measurements of these much smaller capture cross sections
at $E_{0}$ are not anticipated without orders of magnitude
improvements in sensitivity.

For completeness we can also evaluate Eq.~(\ref{eq5}) for the case
of a reaction dominated by a narrow, isolated resonance,

\begin{equation}
 \int\limits_0^\infty {E\,\sigma \left( E \right)}
\,\exp \left( {-E/kT} \right)\,\mathrm{d} E=E_{res} \exp \left( {-E_{res}
/kT} \right)\int\limits_0^\infty \sigma \left( E \right)\,\mathrm{d} E
    \label{eq11}
\end{equation}

Assuming a Breit--Wigner line shape $\sigma \left( E
\right)=\sigma _{BW} \left( E \right)=\pi
\mathchar'26\mkern-10mu\lambda ^2\omega \,\,\frac{\Gamma _1 \Gamma
_2 }{\left( {E_{res} -E} \right)^2+{\Gamma ^2} \mathord{\left/
{\vphantom {{\Gamma ^2} 4}} \right. \kern-\nulldelimiterspace} 4}$
with the proper statistical factor $\omega ={\left( {2j+1}
\right)} \mathord{\left/ {\vphantom {{\left( {2j+1} \right)}
{\left( {\left( {2s_1 +1} \right)\left( {2s_2 +1} \right)}
\right)}}} \right. \kern-\nulldelimiterspace} {\left( {\left(
{2s_1 +1} \right)\left( {2s_2 +1} \right)} \right)}$ yields
\[
\int\limits_0^\infty {\sigma _{BW} \left( E \right)\,\rm{d}E}
=\;\;\frac{\lambda _{res}^2 }{2}\;\omega \;\frac{\Gamma _1 \Gamma
_2 }{\Gamma }\equiv \;\;\frac{\lambda _{res}^2 }{2}\;\omega
\;\gamma .
\]
Expressing $T$ in units of 10$^9$ K and generalizing for a series
of isolated resonances yields

\begin{equation}
 \left\langle {\sigma v} \right\rangle
_{12} =\;\;2.53\times 10^{-13} \mathrm{cm}^3/\mathrm{s} \left( {AT_9 }
\right)^{-3/2} \sum\limits_i {\left( {\omega \gamma } \right)_i \over \mathrm{MeV}}
\times \,\,\exp \left(-E_i /kT  \right).
    \label{eq12}
\end{equation}

For intermediate cases, between the limits of narrow, isolated
resonances and non--resonant reactions, one must directly
integrate Eq.~(\ref{eq5}).

\subsection{Tests of electron shielding}
In the above treatments, it is assumed that the Coulomb potential
of the target nucleus and projectile is that resulting from bare
nuclei. However, for nuclear reactions studied in the laboratory,
the target nuclei and the projectiles are usually in the form of
neutral atoms or molecules and ions, respectively. The electron
clouds surrounding the interacting nuclides act as a screening
potential:~the projectile effectively sees a reduced Coulomb
barrier, both in height and radial extension. This, in turn, leads
to a higher cross section for the screened nuclei, $\sigma_s(E)$,
than would be the case for bare nuclei, $\sigma_b(E)$. There is an
enhancement factor (Assenbaum \textit{et al.} \cite{17})

\begin{equation}
 f_{lab}(E) = \sigma _{s}(E)/\sigma _{b}(E) \sim
\mathrm{exp}(\pi \eta U_{e}/E) \ge 1 \quad \mathrm{for}~~U_{e} << E,
    \label{eq13}
\end{equation}

where $U_e$ is an electron--screening potential energy. This
energy can be calculated, for example, from the difference in
atomic binding energies between the compound atom and the
projectile plus target atoms of the entrance channel. Note that
for a stellar plasma, the value of the bare cross section
$\sigma_b(E)$ must be known because the screening in the plasma
will be quite different from that in the laboratory
nuclear-reaction studies, \textit{i.e.}, $\sigma _{plasma}(E)$ =
$f_{plasma}(E)$ $\sigma_b(E)$, where the plasma enhancement factor
$f_{plasma}(E)$ must be explicitly included for each situation.
The later factor involves the Debye radius of the electrons around
the ions in the plasma, as discussed earlier.  Similarly, a good understanding of
electron--screening effects in the laboratory is needed to extract
a reliable $\sigma_b(E)$ from low-energy data taken with
terrestrial atomic targets. A great deal
can be done experimentally to test our understanding of electron
screening.  Thus, while nuclear reactions are not measured under the
conditions found in stellar interiors, we can test of understanding of the
screening corrections needed to determine $\sigma_b(E)$ from
laboratory data.

Experimental studies of reactions involving light nuclides
(Engstler \textit{et al.} \cite{18}; Strieder \textit{et al.}
\cite{19} and references therein) have shown the expected
exponential screening enhancement of the cross section at low energies. For
example, the cross section of the $^3$He(d,p)$^4$He reaction was
studied over a wide range of energies (Aliotta \textit{et al.}
\cite{20}) where new energy loss information at the relevant low
energies was available for the analysis of the data: the results
led to $U_e$ = 219$\pm $15 eV, significantly larger than the
adiabatic limit from atomic physics, $U_{ad}$ = 119 eV. A weak
point in the analysis is the assumption of the energy dependence
of the bare cross section $\sigma _{b}(E)$. A direct measurement
of $\sigma_b(E)$ using bare nuclides (\textit{e.g.}, a crossed
ion-beam set-up) appears difficult if not impossible due to
luminosity problems. However, a new experimental method for an
indirect $\sigma_b(E)$ determination has been developed, the
Trojan Horse Method (Strieder \textit{et al.} \cite{19} and
references therein). Although it is not yet clear whether all
relevant components of this method have been included in current
analyses, the available data demonstrate how well the indirect
method and the direct measurements complement one another.

There exist various surrogate environments that have allowed
experimentalists to test our understanding of plasma screening effects.
Recently, screening in d(d,p)t has been studied for
deuterated metals, insulators, and semiconductors, \textit{i.e.}
58 samples in total (Raiola \textit{et al.} \cite{21}, Raiola
\textit{et al.} \cite{22} and references therein). As compared to
measurements performed with a gaseous D$_2$ target ($U_e$ = 25
eV), a large effect has been observed in all metals (of order
$U_e$ = 300 eV), while a small (gaseous) effect is found for the
insulators and semiconductors. For the metals, the
hydrogen-solubilities are small (a few percent) leaving the
metallic character of the samples essentially unchanged. An
explanation of the large effects in metals was suggested by the
classical plasma screening of Debye applied to the quasi--free
metallic electrons. The electron Debye radius around the deuterons
in the lattice is given by

\begin{equation}
\label{eq14} R_{D} = (\varepsilon _{o} kT / \alpha n_{eff} \rho
_{a})^{1/2} = 69 (T /n_{eff} \rho _{a})^{1/2}\quad[m]
\end{equation}

\noindent with the temperature of the free electrons $T$ (in units
of K), $n_{eff}$ the number of valence electrons per metallic
atom, and the atomic density $\rho_a$ (in units of m$^{-3})$. With
the Coulomb energy between two deuterons at $R_D$ set equal to
$U_e$, one obtains $U_e$ = (4$\pi \varepsilon _o)^{-1}$
$\alpha$/$R_D$. A comparison of the calculated and observed $U_e$
values determines $n_{eff}$, which is for most metals of the order of
one. The acceleration mechanism of the incident positive ions
leading to the high observed $U_e$ values is thus the Debye
electron cloud at the rather small radius $R_D$, about one tenth
of the Bohr radius. The $n_{eff}$ values have been compared with those
deduced from the known Hall coefficient: within 2 standard
deviations the two quantities agree for most metals. A critical
test of the Debye model is the predicted temperature dependence
$U_e(T) \propto  T^{-1/2}$, which was also confirmed
experimentally. Thus, metals appear to form a ``plasma for the
poor man," allowing terrestrial testing of screening effects very
similar to those found in stellar plasmas.

In the next sections we discuss the present status of the
experimental measurements and extrapolations for each of the
various nuclear reactions in the proton-proton chain and the CN
cycle.

\section{The Proton--Proton Chain and The Carbon--Nitrogen Cycle}
\subsection{The pp and pep reactions}
Unlike the typical stellar nuclear reactions for which rates are
determined by extrapolating laboratory measurements to the needed
$E_{0}$, the weak rates for the pp beta-decay and pep electron
capture reactions are too small to be measured directly in the
laboratory. (More correctly, the Sudbury Neutrino Observatory
experimentalists measured the charged-current breakup reaction on
the deuteron, the inverse of the pp reaction, which has been used
to place some direct constraints on the size of the pp rate.)
Instead the SSM values for these rates are based on model
calculations that take into account the measured values of G$_F \cos{\theta_c}$
and of the axial--vector coupling constant g$_A$ (determined from
the neutron half--life and from superallowed 0$^+\to $0$^+$
beta-decay in nuclei), the nuclear structure governing the
formation of deuterium, exchange current contributions to the weak
amplitude, and weak radiative corrections to that amplitude.

One approach follows in spirit of early estimates of the pp rate
(\textit{e.g.}, Salpeter \cite{23} and Bahcall and May \cite{24}),
and is based on a potential-model description of the deuteron.
Modern treatments employ realistic NN potentials that are derived
from phase-shift analyses of scattering data. Perhaps the primary
nuclear structure uncertainty comes from two--body exchange
currents, which are dominated by short--ranged exchanges involving
poorly known couplings. However these are tightly constrained by
the known strength of tritium beta decay. As the nonrelativistic
two-- and three--nucleon problems can be solved exactly (and
assuming there are no important three-body exchange currents), a
very precise rate estimate can be obtained.

More recently an approach has been taken based on pionless
effective field theory that illustrates nicely the functional
dependence of the pp rate on model--dependent short--range
physics. In these studies \cite{25,26,27} the S--factor is expressed as

\begin{equation}
\hspace*{-0.75 cm} S_{pp} (0) = 3.93 \times 10^{-25} \mbox{MeV} \,
\mbox{b} \; \left(\frac{\Lambda^2}{6.91}\right) \; \left(\frac{G_A
/ G_V}{1.2670}\right)^2 \; \left(\frac{ft(0^+ \to
0^+)}{3073s}\right)^{-1},
    \label{eq15}
\end{equation}

where $\Lambda$ is proportional to the hadronic axial matrix
element connecting the pp and deuteron states, including the
consequence of short--range two--nucleon physics. Effective field
theory determines the functional form of $\Lambda$,

\begin{equation}
\label{eq16} \Lambda = 2.58+0.011\left( {\frac{L_{1,A}
}{1\mbox{fm}^3}} \right)-0.0003\left( {\frac{K_{1,A}
}{1\mbox{fm}^5}} \right)
\end{equation}

where $L_{1,A}$ and $K_{1,A}$ parameterize the two--nucleon
physics at next--to--leading order (NLO) and
next--to--next--to--next--to--leading order (NNNLO), respectively.
As the correction due to $K_{1,A}$ is expected to be much less
than 1$\%$, it follows that the model dependence is embodied in
$L_{1,A}$, a single parameter that determines both S(0) and its
energy dependence. From the work of Schiavilla \textit{et al.}
\cite{28}, this is fixed as 4.2 $\pm$ 2.4 fm$^{3}$ from tritium
beta decay. It follows that

\hspace*{1 cm} $S_{pp}$ (0)=(3.92$\pm$ 0.08)$\times$ 10$^{-25}$
MeVb

The central value is 1$\%$ smaller than that recommended by a 1997
INT working group \cite{29}. The uncertainty encompasses that
earlier result.

A small contribution to pp fusion comes from electron capture on
correlated protons in the solar plasma. To the accuracy required,
the rate for p~+~e$^-$~+~p~$\to$~d~+~$\nu_e$ is proportional to
that for pp beta-decay, $R_{pp}$. Bahcall and May \cite{24} found
\begin{equation}
\label{eq17} R_{pep} =5.51\times 10^{-5}\,\rho \,\,\left( {1+X_H }
\right)T_6^{-1/2} \left( {1+0.02\,T_6 } \right)\,R_{pp}
\end{equation}
where $X_H$ is the mass fraction of hydrogen. In the solar
interior $R_{pep}$ $ \sim $ 2.4 $\times $ 10$^{-3}$ 
$R_{pp}$.

\subsection{The d(p, $\gamma$)$^3$He reaction}
Early measurements (Griffiths \textit{et al.} \cite{30} and Schmid
\textit{et al.} \cite{31}), utilizing D$_{2}$O ice targets sufficiently
thick to stop the incident protons, were used to extract
d(p,$\gamma)$ cross sections down to center--of--mass energies as
low as 10 keV. More recently, using a differentially pumped gas
target in an underground laboratory (LUNA---See Section
\textit{4.3}), Casella \textit{et al.} \cite{32} have measured
the D(p,$\gamma )^{3}$He reaction down to $E_{cm}$ = 2.5 keV, well
below the Gamow peak for this reaction. Across the Gamow peak,
their \textit{measured} data can be characterized by

\hspace*{1 cm} $S$(0) = 0.216 $\pm$~.006 eV--barns

\hspace*{1 cm} $S^{\prime}$(0) = 0.0059 $\pm$~.0004 barns.

The rate of this reaction in the solar interior is so much faster
than that for the pp reaction, that deuterium is effectively
instantaneously converted to $^3$He, and any uncertainty in this
reaction rate has essentially no impact on the pp--chain.

\textit{Other d-burning reactions:}~The d(d,n)$^3$He and d(d,p)$^3$H
reactions have cross--section factors, $S(0)$, which are $\sim
$10$^5$ times larger than for the d(p,$\gamma )^3$He reaction.
However, because the relative deuterium-to-proton abundance ratio
is $\lsim$ 10$^{-17}$ in the solar interior, these other reactions do
not play a significant role. The S-factor for d($^3$He,p)$^4$He
reaction is more than 10$^6$ times larger than that for the
d(p,$\gamma )^3$He reaction, but this is more than counterbalanced
by the small $^3$He/p abundance ratio ($\lsim$ 10$^{-5})$ and the
higher d+$^3$He Coulomb barrier, so that this reaction accounts
for less than 10$^{-3}$ of deuterium burning in the solar interior
(\textit{e.g.}, \cite{33}).

\subsection{The $^3$He($^3$He,2p) $^4$He reaction}
In a main-sequence star burning hydrogen via the pp--chain, the
energy spectrum of the neutrinos emitted from the core is
determined by the relative rates of the various $^3$He--burning
reactions (and the relative burning rates of any subsequent
$^7$Be-burning reactions). This spectrum
determines both the fraction of the pp-chain energy (26.73 MeV) which
is removed from star's energy budget (escaping in the
form of neutrinos)---and the counting rates for various solar neutrino
with differing thresholds and response functions. The rates of the
$^3$He($^3$He,2p)$^4$He, $^3$He($\alpha, \gamma )^7$Be,
$^3$He(d,p)$^4$He, and $^3$He(p,e$^+ \nu_e)^4$He reactions are
discussed below.

Studies conducted in laboratories at the earth's surface of very low 
energy thermonuclear reactions are limited predominantly by
background effects of cosmic rays in the detectors, leading
typically to more than 10 background events per hour. 
Conventional passive or active shielding around the
detectors can only partially reduce the problem of cosmic-ray
background. The best solution is to install an accelerator
facility in a laboratory deep underground. As a pilot project, a
50 kV accelerator facility has been installed in the underground
laboratory at Gran Sasso, where the flux of cosmic--ray muons is
reduced by a factor 10$^6$ compared with the flux at the surface
(Fiorentini \textit{et al.} \cite{34}). This unique project,
called LUNA (Laboratory for Underground
Nuclear Astrophysics), was designed primarily
for a renewed study of the reaction $^{3}$He($^{3}$He,2p)$^{4}$He
at low energies, aiming to reach the solar Gamow peak at $E_{0}\pm
\Delta $/2 = 21$\pm $5 keV. This goal has been reached (Bonetti
\textit{et al.} \cite{35}) with a detected reaction rate of about
1 event per month at the lowest energy, $E$ = 16 keV, with $\sigma $
= 20 fb or 2 $\times$ 10$^{-38}$ cm$^{2}$. Thus, the cross section of an
important fusion reaction of the hydrogen-burning proton-proton
chain has been directly measured for the first time at solar
thermonuclear energies (Fig.~4):~extrapolation is no longer needed
in this reaction. Across the Gamow peak, after correction for
electron shielding (see Eq.~(\ref{eq6}) and its associated text)
their \textit{measured} data can be characterized by \cite{35}

\hspace*{1 cm} $S$(0) = 5.32 $\pm$ 0.08 MeV--barns

\hspace*{1 cm} $S^{\prime}$(0) = -3.7 $\pm$ 0.6 barns.

\subsection{The $^3$He ($\alpha, \gamma$) $^7$Be reaction}
The relative rates of the $^3$He($^3$He,2p)$^4$He and
$^3$He($\alpha, \gamma$)$^7$Be reactions determine the branching
ratio between the ppI and ppII+ppIII terminations of the pp--chain
(see Figure 1). The former produces only the low energy p+p$\to
$d+e$^{+}+\nu _e$ neutrinos ($E_{\nu} \le $ 420 keV), while the
latter produce $^7$Be electron capture neutrinos and the
high--energy $^8$B neutrinos ($E_{\nu}^{max} \sim$ 14.02 MeV), the
neutrinos measured in the Homestake, SNO and Super-Kamiokande 
experiments.

Measurements of the $^3$He($\alpha, \gamma$)$^7$Be reaction (six
via measurements of the direct--capture gamma rays and three via
measurements of the residual $^7$Be activity) were reviewed by a
study group at the INT in 1997 (Section V in Ref. \cite{29}),
with the result that~--

\hspace*{1 cm} $S(0)_{DirectCapture}$= 0.507 $\pm$~.016 keV--b

\hspace*{1 cm} $S(0)_{ResidualActivity}$= 0.572 $\pm$~.026 keV--b

\hspace*{1 cm} $S(0)_{TotalDataSet}$= 0.533 $\pm$~.013 keV--b.

This study group concluded \cite{29} that the grouping of the two
data sets (direct vs.~activity) around two different centroids
``suggests the possible presence of a systematic error in one or
both of the techniques. An approach that gives a somewhat more
conservative evaluation of the uncertainty is to ... determine the
weighted mean of those two results (the direct--capture and
residual-activity data sets). In the absence of information about
the source and magnitude of the excess systematic error, if any,
an arbitrary but standard prescription can be adopted in which the
uncertainties of the means of the two groups (and hence of the
overall mean) are increased by a common factor of 3.7 (in this
case) to make $\chi ^2$=0.46 for one degree of freedom, equivalent
to making the estimator of the weighted population variance equal
to the weighted sample variance. The uncertainty in the
extrapolation is common to all the experiments and is likely to be
only a relatively minor contribution to the overall uncertainty.
The result is our recommended value for an overall weighted mean:

\hspace*{1 cm} $S_{34}$(0) = 0.53 $\pm $ .05 keV--b.''

Now that the $^7$Be(p,$\gamma$)$^8$B reaction rate seems to be on
a firmer footing (see Section \textit{4.5} below), the
$^3$He($\alpha, \gamma$)$^7$Be reaction rate, because of the
inconsistency of the direct--capture and $^7$Be activity results,
has become the most important nuclear uncertainty in the solar
neutrino problem. To try to resolve this issue, at least four new
studies of the $^3$He($\alpha, \gamma$)$^7$Be reaction rate are
currently underway. The approaches include new direct-capture and
activity measurements, as well as a Coulomb-dissociation
experiment. A recent series of $^7$Be--activity measurements
\cite{36} at four energies in the range 420 $<$ $E_{cm} <$ 950 keV
determines an extrapolated intercept of $S_{34}$(0) = 0.53
$\pm$~.02 $\pm$~.01 keV--b.

It should be noted that the $^3$He($\alpha, \gamma$)$^7$Be
reaction exhibits a particularly simple, non--resonant behavior
from its threshold to $E_{cm} $ $\sim$ 2.5 MeV, and its
entrance-channel phase shifts are well determined over this energy
range. Thus its energy dependence provides a good test of nuclear
models. Section V of Ref. \cite{29} provides a good review of
theoretical work on this reaction.

\subsection{The $^3$He(p,e$^+\nu_e$)$^4$He reaction}
The ``hep" reaction proceeds through the same weak--interaction as
the pp reaction (Section \textit{4.1}, above), and as such its
cross section factor would be expected to be orders of magnitude
smaller than for the $^3$He($^3$He,2p)$^4$He and $^3$He($\alpha,
\gamma$)$^7$Be reactions and of no interest as a significant
termination for the pp--chain. However this weak branch produces
energetic neutrinos ($E_{\nu}^{max}$ = 18.77 MeV) that extend
beyond the endpoint of the $^8$B neutrino spectrum, so that in
principle a weak flux could be detected. Both SNO and
Super--Kamiokande have placed interesting limits on this flux;~see
Section 5, below.

The nuclear physics of this reaction is rather subtle:~in a naive
shell--model description, the incident proton would, by the Pauli
exclusion principle, occupy a state orthogonal to the two s--shell
protons in $^3$He. In the allowed approximation, the conversion of
the proton to a neutron by the weak interaction would then produce
a state orthogonal to the na\"{\i}ve s--shell $^4$He ground state.
This suggests that smaller components in the nuclear wave function
and corrections such as exchange currents
might be unusually important in determining the cross section.
For this reason theoretical uncertainties in estimating the rate
are considerably larger than in \textit{pp} beta--decay, despite
modern few-body techniques that can be applied to the continuum
four--body problem \cite{37}. The most recent evaluation of the
rate of this weak-interaction reaction \cite{27} gives an
S--factor of

\hspace*{1 cm} $S_{hep}$(0) = (8.6 $\pm $ 1.3)$\times $10$^{-20}$
keV b,

corresponding to a \textit{hep} termination branch for the
pp--chain of 2.5$\times $10$^{-7}$ and a \textit{hep} neutrino
flux of $\Phi _{hep}$ = 7.9$\times$ 10$^{3}$ cm$^{-2}$ s$^{-1}$ in
BP04 \cite{9}.

\textit{Other ${}^3$He--burning reactions:}~The S--factor $S(0)$ for
$^3$He(d,p)$^4$He is comparable in magnitude to that for
$^3$He($^3$He,2p)$^4$He reaction, while the much smaller relative
abundance of deuterium compared to $^3$He (d/$^3$He $\lsim$
10$^{-12})$ more than compensates for the higher $^3$He+$^3$He
Coulomb barrier. (Deuterium has a very short lifetime $\sim $ 1
sec in the solar core. The $^3$He lifetime increases sharply with
decreasing temperature, but is typically $\sim $10$^6$ years in
the hotter central core of the sun \cite{33}.) The net result is
that the $^3$He(d,p)$^4$He reaction accounts for less than 0.1$\%$
of $^3$He--burning in the solar interior.

\subsection{The $^7$Be (e$^-$, $\nu$)$^7$Li reaction}
The neutral--atom lifetime and the branching ratio (the ratio of
the transition rate to the first excited state $^7$Li$^{\ast }$ to
the total $^7$Li$^{\ast }+^7$Li$_{gs}$ rate) for the
electron--capture decay of $^7$Be is well determined (see Table
7.6 in ref.~\cite{38}):

\hspace*{1 cm} $\tau_{1/2}$ = 53.22 $\pm$ 0.06 days

\hspace*{1 cm} $\omega$($^7$Li$^{\ast}$)/ $\omega$(total)
 = (10.44 $\pm$ 0.06) $\%$.

However, one does not want the neutral--atom lifetime,
but the lifetime for the conditions of the temperature and
density in the solar interior~--~which determine the ionization
state and the electron density governing this electron--capture
decay rate.

Bahcall and Moeller \cite{39} evaluated the rate for capture of a
continuum electron as well as the contribution from bound--state
capture in Be ions that are not totally ionized. While continuum
capture dominates, inclusion of bound--state capture increases the
rate by about 22$\%$. They found

\begin{equation}
    \begin{array}{ll}
 R(^7Be + e^-) = 5.60 \times 10^{-9} (\rho/2) (1 + X_H) T_6^{-1/2}  \\
\hspace*{2.3 cm} \times [1 + 0.004 (T_6 - 16) ] s^{-1}.
    \end{array}
    \label{eq18}
\end{equation}

Gruzinov and Bahcall \cite{40} used the density matrix formalism
(\textit{e.g.}, Feynman \cite{41}) in an alternative calculation
of the rate. The result from this approach, which does not
separate bound--state capture from continuum capture, agrees with
Eq.~(\ref{eq18}) to better than 1$\%$.

\subsection{The $^7$Be (p,$\gamma$)$^8$B reaction}
While the pp--III branch plays only a very minor role (0.013 $\%$)
in solar energy production, the decay of the $^8$B produced in
the$^7$Be(p,$\gamma$) reaction is directly responsible for the
production of essentially all of the solar neutrinos of interest
to the SNO and Super-Kamiokande detectors. Motivated by interest in
quantitatively understanding the solar neutrino fluxes measured in
these detectors, experimentalist have mounted five new direct
measurements of the $^7$Be(p,$\gamma$) reaction in the last decade,
as well as a
number of indirect measurements in which the rate
is deduced either via the strength of a proton--transfer
reaction or via Coulomb dissociation of a $^8$B beam.

Except for the original Kavanagh measurement of the high--energy
positrons \cite{42}, the direct $^{7}$Be(p,$\gamma$) experiments
have measured the reaction cross section by detecting the
delayed alpha-particles, emitted following the $\beta$--decay of
$^8$B ($\tau_{1/2}$ = 0.77 s) to $^8$Be. The five modern, low--energy
($E_{cm}<$ 425 keV) measurements \cite{43,44,45,46,47} are
consistent with each other, and when each is fit with the same
theoretical model (\textit{e.g.}, DB94 \cite{48}) they generate a
weighted average of 21.4 $\pm $ 0.5 eV b \cite{47}. However, there
is an additional uncertainty introduced by the variations between
models that can be fit to the data and then used to perform the
extrapolation to $S_{17}$(0). Junghans \textit{et al.} \cite{47}
compare 12 model calculations (their Fig.~16) all of which cluster
around a central value of $\sim $22 $\pm $ 0.8 eV b. (See also
Jennings \textit{et al.} \cite{49}.) Junghans \textit{et al.}
combine these experimental and theoretical uncertainties as

\hspace*{1 cm} $S_{17}$(0) = 21.4 $\pm $ 0.5 (exp) $\pm $ 0.6
(theory) eV b.

An important systematic uncertainty in these direct
experiments has been the thickness and composition of the $^{7}$Be
targets. To circumvent this issue an experiment is
currently in its final planning stages at ORNL, which will utilize
a $^7$Be beam incident on a hydrogen gas target \cite{50}. (Of
course, this will introduce other, but different,
systematic issues, such as the distribution of the hydrogen gas
in the windowless target and the charge-state distribution of
the recoiling $^8$B nucleus in the recoil separator.)
Two other proposed methods for avoiding the $^7$Be--target
issues are indirect measurements of the reaction rate by
(a) proton--transfer reactions or (b) the Coulomb
dissociation (CD) of accelerated $^8$B ions:

(a) Transfer--reaction experiments \cite{51} have been used to
extract the asymptotic normalization coefficients (ANCs) from the
reactions $^{10}$B($^7$Be,$^8$B)$^9$Be and
$^{14}$N($^7$Be,$^8$B)$^{13}$C coupled with measurements of the
ANCs for the $^{13}$C(p,$\gamma)$ and $^9$Be(p,$\gamma)$
reactions. The ANCs from these reactions determine the behavior of
the tail of the $^7$Be+p wave function in the peripheral region
that dominates capture reactions in such loosely bound nuclei.
This method has been checked by comparing an ANC measurement of
the $^{16}$O($^3$He,d)$^{17}$F reaction with direct
$^{16}$O(p,$\gamma)^{17}$F measurements, yielding agreement at the
level of 9$\%$ \cite{52}. The ANC analysis of the
$^{10}$B($^7$Be,$^8$B)$^9$Be and $^{14}$N($^7$Be,$^8$B)$^{13}$C
reactions by Azhari \textit{et al.} \cite{51} gave
$S_{17}$(0) = 17.3 $\pm$ 1.8 eV b. An ANC analysis \cite{53} of
a number of $^{8}$B break--up studies (involving a variety of
energies and targets) yielded a value of $S_{17}$(0) = 17.4 $\pm $
1.5 eV b.  However, it should be noted that a different, separate
analysis \cite{54,55} of the $^{12}$C($^8$B,$^7$Be)$^{13}$N
break--up reaction gave $S_{17}$(0) = 21.2 $\pm $ 1.3 eV b, indicating
that substantial uncertainties remain in the interpretation of the measurements.

(b) Several $^8$B Coulomb--dissociation experiments (in which the
break--up products are measured at very forward angles,
corresponding to very large impact parameters and hence very small
nuclear contributions) have been carried out for $^8$B beam
energies from 52 to 254 MeV/nucleon. The four most recent
Coulomb dissociation experiments \cite{56,57,58,59} deduce
$S_{17}$(0) values in the range from 17.8 to 20.6 eV b. In the
absence of a separate calibration of the CD method, at this time
the comparison of the these $^8$B CD measurements with the direct
$^7$Be(p,$\gamma)$ cross section measurements may
be more important as a
check on the level of validity of the CD method for determining
(p,$\gamma)$ S--factors. Such a check is important in assessing CD
for other cases where direct measurements may not be possible.

\subsection{The Carbon--Nitrogen Cycle}
Hydrogen burning via the CN--cycle (Fig.~2) accounts for less than
2$\%$ of the $^4$He produced in the sun. In their
analysis of the status of the various nuclear reactions in this
cycle, the 1997 INT study group \cite{29} concluded that by far
the most significant uncertainty in this cycle was the poorly
defined role of the subthreshold resonance ($E_{cm}$ = -504 keV)
in the extrapolation of the measured rate of the
$^{14}$N(p,$\gamma)^{15}$O reaction. This reaction is $\sim$100
times slower than the other reactions in the CN--cycle and
therefore determines the overall rate of this cycle. Based on
their analysis of the data of Lamb $\&$ Hester \cite{60} and
Schr\"{o}der \textit{et al.} \cite{61}, the INT workshop participants recommended
$S_{tot}$(0) = $3.5_{-2.0}^{+1.0}$ keV b, and emphasized the
need for new experiments to improve our understanding of the
$^{14}$N(p,$\gamma)^{15}$O reaction. Subsequently, in a further
R--matrix reanalysis of the Schr\"{o}der data, Angulo and
Descouvement \cite{62} extracted a reduced proton width for the
subthreshold resonance corresponding to the 6.79--MeV state that
was much smaller than the one used in the original Schr\"{o}der
analysis. This smaller reduced proton width resulted in a value of
$S_{tot}$(0) = 1.77 $\pm$~0.20 keV b for this reaction \cite{62}.

Recently, preliminary results have become available from two new,
independent measurements of the rates for each of the gamma--ray
transitions for the $^{14}$N(p,$\gamma)^{15}$O reaction, at the
LENA (UNC--TUNL) facility and at the LUNA (Bochum--GranSasso)
facility. The LENA measurements \cite{63} utilized an implanted
target ($^{14}$N atoms implanted in 0.5--mm thick Ta backings) and
were carried out down to $E_{cm}$ = 145 keV. The LUNA measurements
\cite{64} utilized both a solid target (for $E_{cm}\ge$ 130 keV)
and a differentially--pumped N$_2$ gas target (for $E_{cm} \ge$ 70
keV). These LENA and LUNA data sets are in good agreement, well
within their statistical uncertainties, and when combined with a
remeasurement \cite{65} of the lifetime of the subthreshold state
at $E_x$ = 6.79 MeV, these new measurements indicate a factor of
$\sim$2 reduction in the $S_{tot}$(0) value for this reaction, to
1.67 keV b, compared to the Schr\"{o}der value of 3.5 keV b
\cite{61,29}. (Preliminary reports for these two data sets
indicate statistical uncertainties of $\sim$ 5$\%$ and systematic
uncertainties of $\sim$ 9$\%$ in their determination of
$S_{tot}$(0). However, it should be noted that the analysis of the
70--keV LUNA data point is not yet complete.)

A preliminary examination \cite{63,64} of the consequences of this
factor of $\sim$2 reduction in the rate of the
$^{14}$N(p,$\gamma)^{15}$O reaction finds (a) a corresponding
factor of $\sim$2 reduction in the CN component of the solar
neutrino flux and (b) an increase of $\sim$1 Gyr in globular
cluster ages based on turn--off points from the main--sequence.  When
the data are final, the effects on detailed SSM neutrino flux predictions
can be evaluated more quantitatively.

\section{Solar Neutrino Fluxes:~Theory and Experiment}
The nuclear cross sections discussed here, together with other
solar model parameters (\textit{e.g.}, abundances, the solar age),
can be incorporated into solar model calculations, leading to
predictions that can be verified in helioseismology or neutrino
flux measurements. The principal neutrino fluxes from the pp chain
are the pp and $^8$B beta--decay sources and the line sources
(broadened by about 2 keV from thermal effects) from electron
capture on $^7$Be. As seen from Figure 1, one of the reasons for
the importance of these fluxes is that they provide a direct test
of the competition among the ppI, ppII, and ppIII cycles that
comprise the pp chain:~the $^7$Be and $^8$B neutrinos tag the ppII
and ppIII cycles, respectively, while the pp neutrinos then
determine the total rate of fusion in the sun. This competition is
an important check on the SSM, specifically its prediction of the
sun's central core $T_c$. For example, the $^8$B flux is found to
vary as $\sim$ $T_c^{24}$ when solar model input parameters are
varied (while conserving the luminosity), while the flux ratio
$\phi (^7$Be)/$\phi (^8$B) $\sim $ $T_c^{-12}$. The precision with
which these fluxes can be predicted in the SSM is also crucial to
current efforts to better determine the parameters governing solar
neutrino oscillations.

Results from the most recent Bahcall and Pinsonneault SSM (BP04)
\cite{9} are summarized below, where E10 = 10$^{10}$:

\begin{table}[htbp]
\begin{center}
\begin{tabular}{|p{46pt}|l|l|}
\hline \textit{Source}& \textit{BP04 (cm}$^{-2} s^{-1})$&
\textit{BP04+ (cm}$^{-2} s^{-1})$ \\
\hline pp& 5.94(1$\pm $0.01) E10&
5.99 E10 \\
\hline pep& 1.40(1$\pm $0.02) E8&
1.42 E8 \\
\hline hep& 7.88(1$\pm $0.16) E3&
8.04 E3 \\
\hline $^{7}$Be& 4.86(1$\pm $0.12) E9&
4.65 E9 \\
\hline $^{8}$B& 5.79(1$\pm $0.23) E6&
5.26 E6 \\
\hline $^{13}$N& 5.71(1$\pm $0.36) E8&
4.06 E8 \\
\hline $^{15}$O& 5.03(1$\pm $0.41) E8&
3.54 E8 \\
\hline $^{17}$F& 5.91(1$\pm $0.44) E6&
3.97 E6 \\
\hline
\end{tabular}
\label{tab1}
\end{center}
\end{table}

These results reflect the dominance of the ppI cycle in the
pp--chain for present--day solar conditions: $\sim$ 85$\%$ of the
sun's energy in generated via this path. The ppII ($\sim$15$\%$)
and ppIII ($\sim $0.02$\%$) account for the remainder.
Approximately 1.7$\%$ of $^4$He synthesis occurs via the CNO
cycle. Cross section uncertainties lead to substantial ranges in
the predicted fluxes.

The cross section uncertainties discussed in this paper are
included in the overall model uncertainties quoted for BP04, which
uses older element abundances from Grevesse and Sauval \cite{66}.
Also shown are the results for BP04+, a calculation identical to
BP04, but using more recent analyses of the solar surface
abundances (C, N, O, Ne, Ar) of Allende Prieto \textit{et al.}
\cite{67}. Contributing to the uncertainties in extracting solar
surface abundances are systematics due to blending of atomic lines
and uncertainties in modelling the solar atmosphere. The Allende
Prieto \textit{et al.} abundances substantially lower the surface
heavy element to hydrogen ratio (from 0.0229 to 0.0176), which
leads to a shallower convective zone and discrepancies with
helioseismology. The BP04+ neutrino fluxes reflect a shift in the
pp chain toward the ppI cycle because the lower metalicity
produces a somewhat cooler sun. Because of its temperature
sensitivity, the $^8$B flux shows the largest change among pp
fluxes, a 10$\%$ decrease. The lower abundances for C, N, and O
directly influence CN--cycle neutrino fluxes, as does the lower
value for $T_c$, leading to changes on the order of 40$\%$. BP04
neutrino spectrum is shown in Fig.~5.  The $^8$B neutrino spectrum
deviates slightly from an allowed shape
because the final $^8$Be state is a broad resonance.

The prospect of directly probing the thermonuclear reactions
occurring in the solar core helped to motivate Ray Davis and his
colleagues \cite{68} to mount their pioneering chlorine solar
neutrino experiment. Sited nearly a mile underground in the
Homestake Mine to avoid cosmic ray--induced backgrounds, this
radiochemical detector operated almost continuously from its
construction in 1967 until 2002. The experiment exploited a
reaction, $^{37}$Cl($\nu _e$,e$^-)^{37}$Ar, that had been first
suggested by Pontecorvo \cite{69} and by Alvarez \cite{70}. The
technique involved the collection of the product $^{37}$Ar
($\tau_{1/2} \sim$ 35 d) from a tank containing 0.61 kiloton of
perchloroethylene (C$_2$Cl$_4)$. Gas was removed from the tank about
every two months by a helium purge, then circulated
through a condenser, a molecular sieve, and a charcoal trap cooled
to the temperature of liquid nitrogen. The resulting efficiency
for collecting the $^{37}$Ar was high, typically greater than
95$\%$, as was demonstrated by the recovery of known amounts of
the carrier gases $^{36}$Ar or $^{38}$Ar that were introduced at
the start of each run. After extraction, the trap was heated and
swept by helium. The extracted gas was passed through a hot
titanium filter to remove reactive gases, and the argon was then
separated from other noble gases by gas chromatography. The
purified argon was loaded into a miniaturized gas proportional
counter, where the subsequent electron capture on $^{37}$Ar was
measured through the 2.82 keV of energy that is released as the
atomic electrons in $^{37}$Cl adjust to fill the K--shell vacancy.
The counting typically continued for about a year ($\sim$ 10 half
lives).

After accounting for small contributions from neutron-- and
comic--ray--induced backgrounds, Davis determined a solar neutrino
production rate of $\sim$ 0.5 $^{37}$Ar atoms/day in the tank.
Despite the low counting rate, the final results for the
experiment achieved an impressive accuracy, 2.56 $\pm $ 0.16
(\textit{stat}) $\pm$ 0.16 (\textit{syst}) SNU, where a Solar
Neutrino Unit (SNU) is 10$^{-36 }$captures/$^{37}$Cl atom/s
\cite{71}. The chlorine experiment is primarily sensitive to $^8$B
neutrinos ($\sim$ 76$\%$), which can induce the strong
super--allowed transition to the 4.99 MeV state in $^{37}$Ar, and
$^7$Be neutrinos ($\sim$16$\%$), which are sufficiently energetic
to overcome the 814 keV threshold of the ground--state transition.
The discrepancy between the chlorine results and the SSM
predictions (8.5 $\pm$ 1.8 SNU for BP04) was the genesis of the
solar neutrino problem.

Two similar radiochemical experiments, SAGE and GALLEX, began
solar neutrino measurements in January 1990 and May 1991,
respectively. These exploited the reaction
$^{71}$Ga($\nu_e$,e$^-)^{71}$Ge which, because of the strength and
low Q--value for the ground--state transition, is primarily
sensitive to the low--energy pp neutrinos. SAGE, which employed 60
tons of liquid Ga metal, continues to operate in the Baksan
Neutrino Observatory, under 4700 mwe of shielding provided by
Mount Andyrchi in the Caucasus. GALLEX and its successor GNO,
which used 30.3 tons of Ga as GaCl$_3$ in a hydrochloric acid
solution, operated until very recently in the Gran Sasso
Laboratory in Italy (3300 mwe).

The primary challenge in both experiments is the more complicated
chemistry of Ge extraction, which is done about every three weeks.
In SAGE the Ge is separated by vigorously mixing into the gallium
a mixture of hydrogen peroxide and dilute hydrochloric acid. This
produces an emulsion, with the Ge migrating to the surface of the
emulsion droplets where it is oxidized and dissolved by
hydrochloric acid. The Ge is extracted as GeCl$_4$, purified and
concentrated, synthesized into GeH$_4$, and further purified by
gas chromatography. The overall efficiency, determined by
introducing a Ge carrier, is typically 80$\%$. In GALLEX/GNO the
Ge is recovered as GeCl$_4$ by bubbling nitrogen through the
solution and then scrubbing the gas through a water absorber. The
Ge is further concentrated and purified, and finally converted
into GeH$_4$. The overall extraction efficiency is typically
99$\%$.

In both experiments the GeH$_4$ is inserted into miniaturized gas
proportional counters, carefully designed for radiopurity, and the
$^{71}$Ge is counted as it decays back to $^{71}$Ga ($\tau_{1/2}$
= 11.43 d). As in the case of $^{37}$Ar, the only signal is the
energy deposited by Auger electrons and X rays that accompany
atomic rearrangement in Ga. Both K and L captures can be detected.
Intense $^{51}$Cr neutrino sources were used to calibrate both
detectors \cite{72}. SAGE is currently conducting a test
calibration with a $^{37}$Ar source, in preparation for an
anticipated repeat of this experiment with $\sim$ 2 MCi of
$^{37}$Ar \cite{73}.

The most recent results of SAGE give a counting rate of 66.9 $\pm$
3.9 (\textit{stat}) $\pm$ 3.6 (\textit{syst}) SNU \cite{74}, while
the GALLEX/GNO result is 69.3 $\pm$ 4.1 (\textit{stat}) $\pm$ 3.6
(\textit{syst}) SNU \cite{75}. The rates are in good agreement,
but well below the SSM prediction (131 $\pm$ 11 SNU in BP04).

Two types of direct--counting experiments have also been done,
using the water Cerenkov detectors Kamiokande and
Super--Kamiokande and the heavy-water detector SNO (Sudbury
Neutrino Observatory). Kamiokande, a 4.5--kiloton cylindrical
imaging water Cerenkov detector, was originally designed for
proton decay but was later reinstrumented to detect $^8$B solar
neutrinos. (The improvements included sealing the detector against
radon inleakage and recirculating the water through ion exchange
columns.) It operated for nearly a decade (1987--95) detecting
solar neutrinos by the Cerenkov light produced by recoiling
electrons in the reaction $\nu_x$ + e $\to \nu_x^\prime$ +
e$^\prime$. Both electron and heavy--flavor neutrinos contribute
to the scattering, with $\sigma (\nu_e)$/$\sigma (\nu _{\mu })
\sim$ 7. The threshold of the detector was initially 9.5 MeV but
was ultimately lowered to 7 MeV, due to Kamiokande III
improvements in electronics and the addition of wavelength
shifters to improve PMT light collection. The inner 2.14 kilotons
of water was viewed by 948 50--cm photomultiplier tubes (PMTs),
which provided 20$\%$ photocathode coverage. An additional 123
PMTs viewed the surrounding 1.5m of water, which served as an
anticounter. The fiducial volume employed for solar neutrino
measurements was the central 0.68 kilotons, the region most
isolated from the high--energy gamma rays generated in the
surrounding rock walls of the Kamioka mine. Kamiokande found a
$^8$B neutrino flux of (2.80 $\pm$ 0.19 $\pm$ 0.33) $\times$
10$^6$/cm$^2$s \cite{76}, 48$\%$ of the BP04 SSM prediction of
(5.79 $\pm$ 1.33) $\times$ 10$^6$/cm$^2$s.

The experiment was remarkable in several respects. It was the
first to measure solar neutrinos in real time. It exploited the
sharp forward peaking of the scattered electrons, in the direction
of the incident neutrino, to extract solar neutrino events from a
significant but isotropic background. The ``pointing'' of events
back to the sun provided the first direct evidence that neutrinos
originated from the sun.

Super--Kamiokande, the massive 50--kton successor to Kamiokande,
collected 1496 days of solar neutrino data over its first run,
from May 1996 through July 2001. With its high PMT coverage
(40$\%$), lower threshold ($\sim$ 5 MeV), and much larger fiducial
volume (22.5 ktons), Super--Kamiokande collected data at a rate
$\sim$ 100 times faster than that of its predecessor, Kamiokande.
The result was an extraordinary set of precise measurements of the
flux, recoil electron spectrum, and seasonal (earth's orbital
eccentricity) and day/night (earth regeneration of neutrinos) time
variations of the solar neutrinos. The experiment placed a
significant constraint on the hep flux ($<$ 7.3 $\times$
10$^4$/cm$^2$s, a limit about 10 times the BP04 prediction) and on
various exotic neutrino properties, such as magnetic moments.
Super--Kamiokande's $^8$B neutrino flux result is (2.35 $\pm$ 0.19
$\pm$ 0.33) $\times$ 10$^6$/cm$^2$s \cite{77}, 41$\%$ of the BP04
prediction.

The SNO detector was built deep in the Creighton $\#$9 nickel mine
in Sudbury, Ontario, for the purpose of determining the flavor
content of the $^8$B solar neutrino flux. A central acrylic vessel
containing one kiloton of heavy water is surrounded by 5m (7
ktons) of light water to shield the inner volume from neutrons and
gammas. The detector is viewed by 9500 20--cm PMTs, providing
56$\%$ coverage.

The heavy water allows the experimenters to exploit three
reactions with varying flavor sensitivities

\hspace*{1 cm} $\nu _e$ + d $\to $ p + p + e$^-$~~(CC:~charged
current)

\hspace*{1 cm} $\nu _x$ + d $\to \nu_x^\prime$ + n +
p~~~(NC:~neutral current)

\hspace*{1 cm} $\nu _x$ + e$^-$ $\to \nu_x^\prime$ +
e$^-$$^\prime$~~~~~(ES:~elastic scattering)

As the Gamow--Teller threshold for the CC reaction is concentrated
close to the p + p threshold of 1.44 MeV, most of the neutrino
energy is transferred to the electron. Thus the CC electron
spectrum provides a better probe of $\nu_e$ spectral distortions
than does the ES electron spectrum.

The NC reaction, which is observed through the produced neutron,
provides no spectral information, but does measure the total solar
neutrino flux, independent of flavor. The SNO experiment has used
two techniques for measuring the neutrons. In the initial SNO pure
D$_2$O phase the neutron signal was capture on deuterium, which
produces 6.25 MeV gammas. In a second phase 2.7 tons of salt was
added to the heavy water so that Cl would be present to enhance
the capture, producing 8.6 MeV gammas. The NC and CC events can be
separated reasonably well because of the modest backward peaking
($\sim$1~--~cos($\theta)$/3) in the angular distribution of the
latter. This allowed the experimenters to determine the total and
electron neutrino fraction of the solar neutrino flux. A third
phase has begun in which direct neutron detection is provided in
pure D$_2$O by an array of $^3$He--filled proportional counters.

The ES, which we noted previously measures electron and
heavy-flavor neutrinos, but with reduced sensitivity to the
latter, is the same reaction exploited by Super--Kamiokande.
Because the SNO volume is smaller, the SNO ES statistics are
reduced. However, the lower SNO backgrounds lead to a smaller
systematic error, so that the overall the SNO and
Super--Kamiokande sensitivities are comparable. Because the ES and
CC scattered electrons are measured in the same detector, the ES
reaction provides an important cross check on the consistency of
the results from the CC and NC channel.

The SNO results are presented in Fig.~6. The deduced fluxes are
\cite{78}

\hspace*{1 cm} $\phi_{CC}$ = (1.59 $\pm$ 0.08 $\pm$ 0.07) $\times$
10$^6$/cm$^2$s

\hspace*{1 cm} $\phi_{ES}$ = (2.21 $\pm$ 0.28 $\pm$ 0.10) $\times$
10$^6$/cm$^2$s

\hspace*{1 cm} $\phi _{NC}$ = (5.21 $\pm$ 0.27 $\pm$ 0.38)
$\times$ 10$^6$/cm$^2$s

The total flux (NC) is in agreement with the SSM prediction, but
the CC results indicate that almost 70$\%$ of the solar neutrinos
arrive on earth as heavy--flavor neutrinos. The ES results are in
excellent accord with this conclusion, as Fig.~6 shows, as well
as with the Super-Kamiokande results.  The
data rule out the hypothesis of no flavor change at greater
than 7$\sigma$.

\section{Conclusions}

This paper has reviewed two problems. The first was the 70--year
effort to understand solar energy generation and stellar evolution
as a consequence of the thermonuclear reactions occurring in the
interiors of stars. It led to collaborations between solar
modelers, who constructed codes to describe the evolution of our
sun and to predict its helioseismology and neutrino production,
and laboratory experimentalists, who provided the nuclear reaction
data essential to the SSM. The second was the nearly 40--year
program to measure the flux, flavor, and spectrum of solar
neutrinos. This effort began with the pioneering experiment of Ray
Davis, Jr.~and his collaborators at Homestake, and led to the
current generation of sophisticated active detectors, SNO and
Super--Kamiokande.

Both programs required great patience---years of effort to refine
the SSM and to develop better solar neutrino detection
techniques---before it became clear that the resolution of the
solar neutrino problem had to be new physics. That resolution
included a new phenomenon, matter enhancement of neutrino
oscillations, which altered our view of neutrino oscillations and
opened up new experimental tests, such as spectral distortions and
day--night effects. This further motivated the development of
robust detectors like SNO and Super--Kamiokande.

The conclusion that was reached after decades of work is quite
profound:~the solar neutrino discrepancy first identified by Davis
is due to new particle physics, neutrino oscillations. This
phenomenon lies outside the minimal standard model, requiring
massive neutrinos and a nontrivial relationship between mass and
weak--interaction eigenstates. Analyses that take into account not
only the solar neutrino data described above, but also the KamLAND
reactor neutrino data \cite{79}, conclude \cite{80}

\hspace*{1 cm} $\delta$m$_{21}^2$ = m$_2^2$-m$_1^2 \quad \sim$
(8.2 $\pm$ 0.3 $\pm$ 0.9) $\times$ 10$^{-5}$ eV$^2$

\hspace*{1 cm} tan$^2\theta_{12} \quad \sim$ 0.39 $\pm$ 0.05 $\pm$
0.15,

\noindent where $\theta_{12}$ is the neutrino mixing angle and m$_1$
and m$_2$ the mass eigenvalues.  The
KamLAND data are very important in further constraining the mass
difference and thus the vacuum oscillation length.

This explanation of the solar neutrino puzzle raises new
questions. What accounts for the scale of neutrino masses, a scale
so different from other standard-model fermion masses? Why are
neutrino mixing angles large? (The corresponding
atmospheric--neutrino mixing angle $\theta_{23}$ is maximal, to
within measurement uncertainties.) There are many parameters in
the neutrino mixing matrix not yet determined. These include the
absolute scale of neutrino mass, the hierarchy (where the nearly
degenerate pair of neutrinos that participate in solar neutrino
oscillations are lighter than or heavier than the third neutrino),
the size of the third mixing angle $\theta_{13}$, the charge
conjugation properties of neutrinos (that is, whether the masses
are generated from Majorana or Dirac mass terms), and the size of
the CP--violating phases that appear in the mixing matrix. The
answers to these questions could help point the way to important
extensions of the standard model.

One of the most remarkable outcomes of the solar (and atmospheric)
neutrino problems is that the neutrino mass differences that
emerged are in a range accessible to terrestrial experiments with
accelerator and reactor neutrinos, such as KamLAND. Despite
the importance of future terrestrial experiments---such as double
beta decay searches for Majorana masses and long-baseline neutrino
experiments to look for CP violation---solar neutrinos still have
an important role to play. For the foreseeable future
(\textit{e.g.}, until beta beams are developed), solar neutrinos
will remain our only intense source of electron neutrinos. Some
future goals that might be reached by exploiting these neutrinos
include \cite{81}:

* \textit{A low--energy solar neutrino measurement to test the
consequences of a large $\theta_{12}.$}  One feature of the
accepted LMA (Large Mixing Angle) neutrino oscillation explanation
of current data is that the survival probability for low energy
neutrinos is substantially higher than that for the $^8$B
electron neutrinos measured in SNO's
CC channel. Verifying this prediction is important.

* \textit{Future high--statistics pp solar neutrino experiments to
improve our knowledge of $\theta_{12}.$}  Solar neutrino
experiments are the primary source of information on
$\delta$m$_{12}^2$ and $\theta_{12}$. While KamLAND has helped to
narrow the LMA solution region in $\delta$m$_{12}^2$, neither it
nor Borexino will appreciably improve our knowledge of the mixing
angle. A pp solar neutrino measurement with an uncertainty of
3$\%$ would constrain $\theta_{12}$ with an accuracy comparable to
that possible with the entire existing set of solar neutrino data.
Thus a goal of next--generation pp solar neutrino experiments is
to achieve 1$\%$ uncertainty, thus substantially tightening the
constraints on $\theta_{12}$.

* \textit{Precise flux measurements of the low--energy
(pp and}$^7$\textit{Be) solar neutrinos to improve limits on
sterile neutrinos coupling to the electron neutrino.} Sterile
neutrinos with even small couplings to active species can have
profound cosmological effects. The coupling of $\nu_e$s to sterile
states can be limited by CC and NC measurements with an accurately
known neutrino source. KamLAND (together with the SNO CC and NC
data) should ultimately limit the sterile component of $^8$B
neutrinos to $\sim$ 13$\%$. This bound can be improved by
measuring the NC and CC interactions of pp and $^7$Be neutrinos to
accuracies of a few percent. In general one expects the sterile
component of solar neutrino fluxes to be energy dependent. Thus
low--energy solar neutrino experiments are an important part of
such searches.

* \textit{Preliminary estimates indicate that limits on the
neutrino magnetic moment could be improved by an order of
magnitude in future pp neutrino ES experiments.} A neutrino
magnetic moment will generate an electromagnetic contribution to
neutrino ES, distorting the spectrum of scattered electrons. The
effect, relative to the usual weak amplitude, is larger at lower
energies, making a high-precision pp solar neutrino experiment an
attractive testing ground for the magnetic moment. The strongest
existing laboratory limit, $\mu (\nu_e) <$ 1.5 $\times$ 10$^{-10}$
Bohr magnetons, could be improved by an order of magnitude in a
1$\%$ experiment.

* \textit{Future pp solar neutrino experiments to probe CPT
violation with an order of magnitude more sensitivity than has
been achieved in the neutral kaon system.} As neutrinos are
chargeless, they provide an important testing ground for CPT
violation. If CPT is violated, the neutrino mass scale and mass
splittings will differ between neutrinos and antineutrinos. (This
has been offered as a possible explanation of the LSND
neutrino oscillation results.) A
high precision measurement of the pp $\nu_e$survival probability
is an important component of CPT violation searches. In some
models this will test CPT violation at a scale $<$ 10$^{-20}$ GeV,
which can be compared to the present bound from the neutral kaon
system, $<$ 4.4 $\times$ 10$^{-19}$ GeV.

* \textit{Solar neutrino detectors are superb supernova detectors.}
Many of the most interesting features in the supernova neutrino
``light curve" are flavor specific and occur at late times, 10 or
more seconds after core collapse. Because of their low thresholds,
flavor specificity, large masses, and low backgrounds, solar
neutrino detectors are ideal for following the neutrino emission
out to late times. Solar neutrino detectors will likely be the
only detectors capable of isolating the supernova $\nu_e$ flux
during the next supernova. The 3 ms deleptonization burst,
important in kinematic tests of neutrino mass, is mostly of this
flavor. Electron--flavor neutrinos also control the isospin of the
nucleon gas---the so called ``hot bubble"---that expands off the
neutron star. The hot bubble is the likely site of the r--process.
Currently we lack a robust model of supernova explosions. The open
questions include the nature of the explosion mechanism, the
possibility of kaon or other phase transitions in the high density
protoneutron star matter, the effects of mixed phases on neutrino
opacities and cooling, possible signatures of such phenomena in
the neutrino light curve, and signals for black hole formation.
Supernovae are ideal laboratories for neutrino oscillation
studies. One expects an MSW crossing governed by $\theta_{13}$,
opening opportunities to probe this unknown mixing angle. The MSW
potential of a supernova is different from any we have explored
thus far because of neutrino--neutrino scattering contributions.
Oscillation effects can be unravelled because the different
neutrino flavors have somewhat different average temperatures. All
of this makes studies of supernova neutrino arrival times, energy
and time spectra, and flavor composition critically important.
Finally, the detection of the neutrino burst from a galactic
supernova will provide an early warning to optical
astronomers:~the shock wave takes from hours to a day to reach the
star's surface.

\textit{* Future solar neutrino experiments could provide crucial
tests of the SSM.} The delicate competition between the ppI, ppII,
and ppIII cycles comprising the pp chain is sensitive to many
details of solar physics, including the core temperature, the
sun's radial temperature profile, the opacity, and the metalicity.
We have noted that this competition can be probed experimentally
by measuring the $^7$Be flux (which tags the ppII cycle), the
$^8$B flux (which tags the ppIII cycle), and the pp flux (which
effectively tags the sum of ppI, ppII, and ppIII). Precise
measurements of the pp and $^7$Be neutrino fluxes, combined with
further improvements in nuclear cross section determinations,
could significantly improve our understanding of current
conditions in the solar core.

$\ast$ \textit{The CN neutrino flux is an important test of
stellar evolution.} The CN cycle is thought to control the early
evolution of the sun, as out--of--equilibrium burning of C, N, and
O powers an initial convective solar stage, thought to last about
10$^8$ years. Furthermore, one of the key SSM assumptions equates
the initial core metalicity to today's surface abundances. A
measurement of CN cycle neutrinos would quantitatively test this
assumption. The recent controversy about surface elemental
abundances provides additional motivation.

This work was supported in part by the US Department of Energy
under grants DE-FG02-00ER-41132, DE-FC02-01ER-41187 (SciDAC), and
DE-FG02-91ER-40609.

\newpage

\begin{figure}
\includegraphics{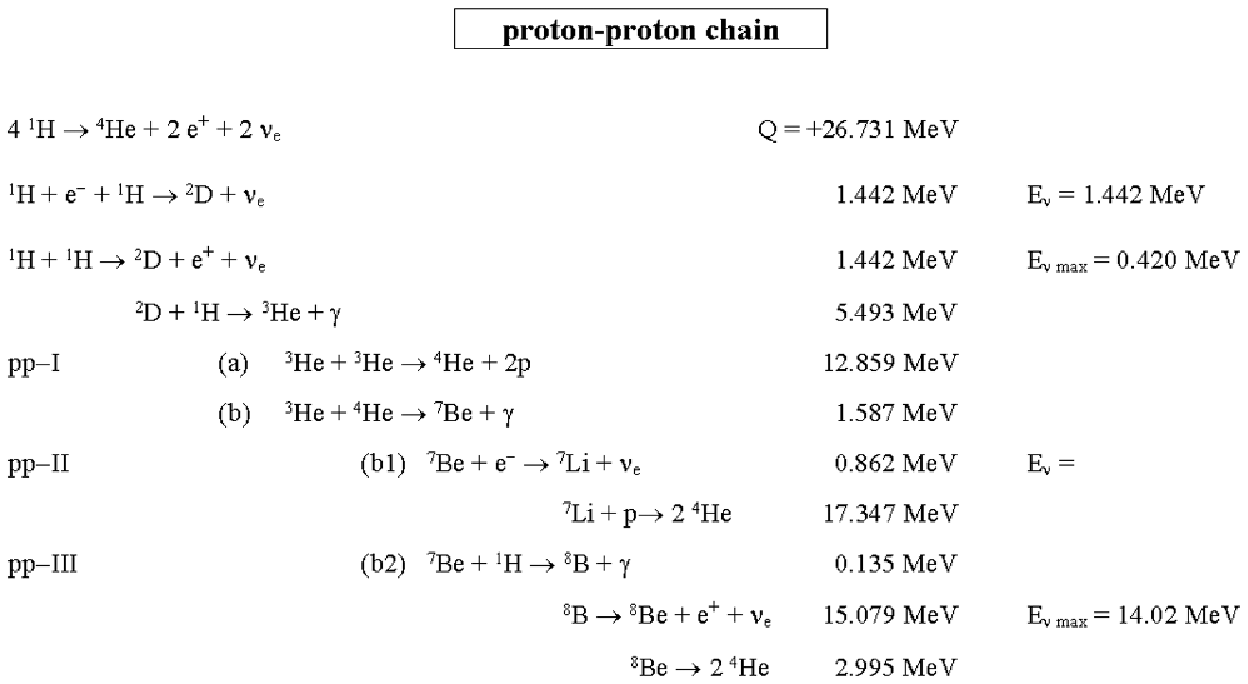}
\caption{The pp--chain, showing the three terminations:~ppI,
ppII, and ppIII.}
\end{figure}

\begin{figure}.
\includegraphics{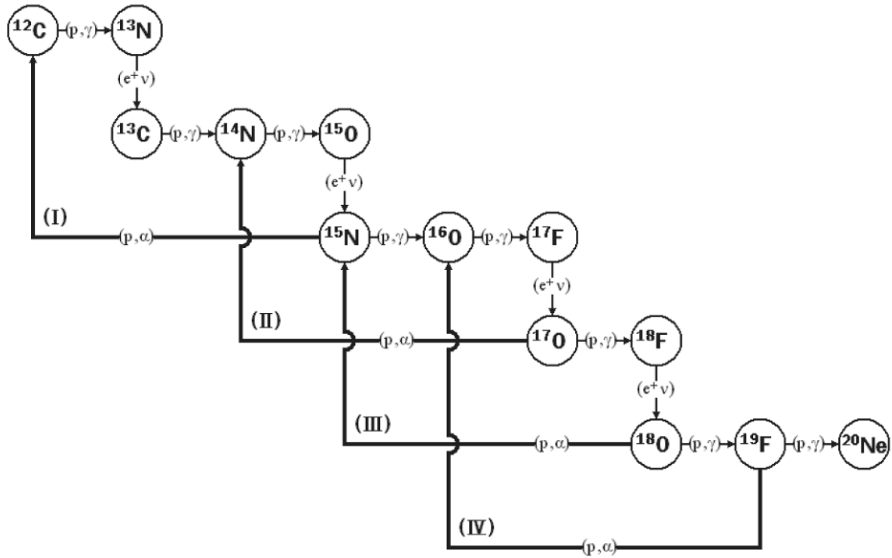}
\caption{The primary CN--cycle, together with three subsidiary
cycles which can bypass the $^{15}$N(p,$\alpha)^{12}$C termination
to extend this processing as far as $^{20}$Ne.}
\end{figure}

\begin{figure}
\includegraphics{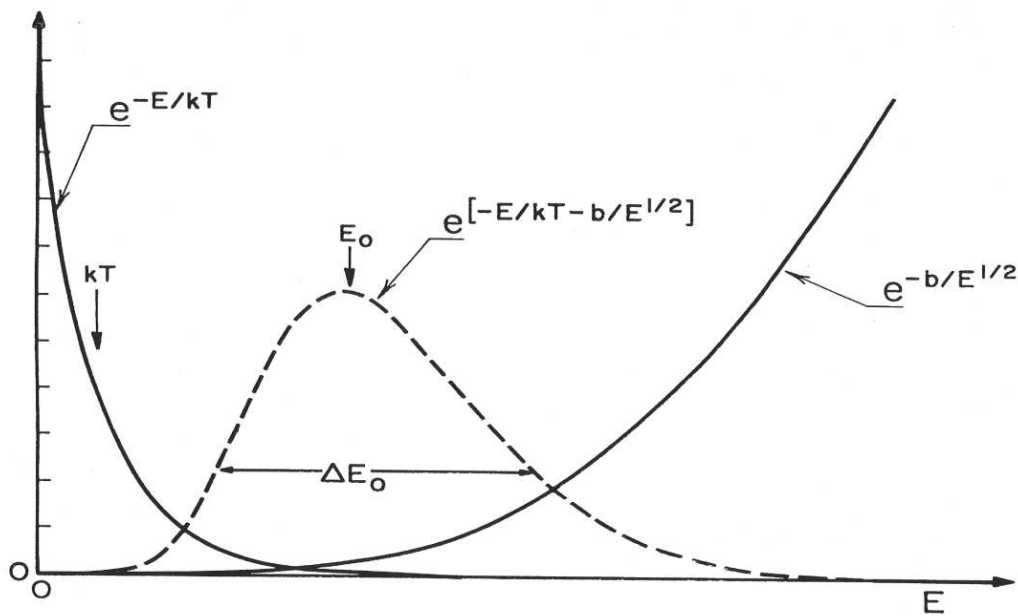}
\caption{The energy dependence of the integrand in Eq.
(\ref{eq6}), defining the parameters of the Gamow peak.}
\end{figure}

\begin{figure}
\includegraphics{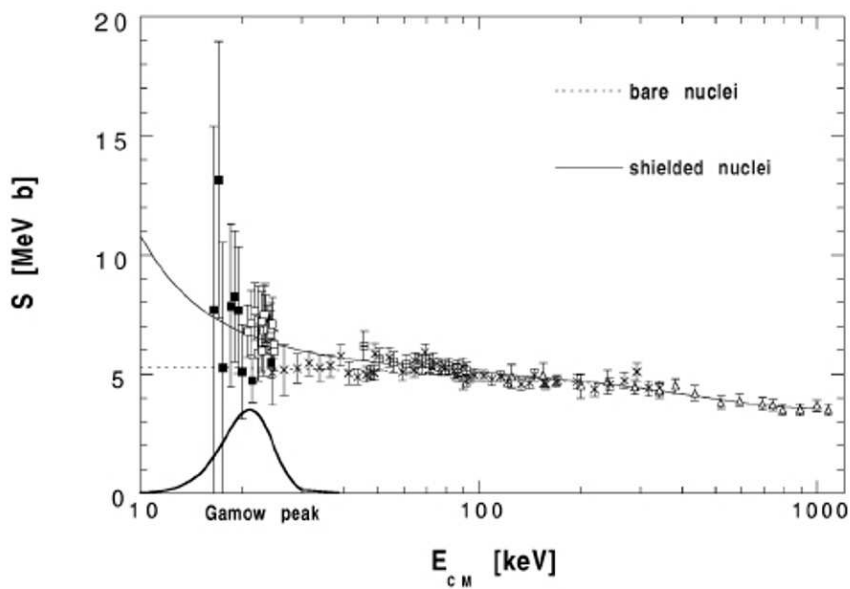}
\caption{The measured energy dependence of the astrophysical S(E)
factor for the $^3$He($^3$He,2p)$^4$He reaction \cite{35}. The
solid curve is a fit to the data including the electron shielding
parameters deduced by Aliotta \textit{et al.} \cite{20} from a
study of the $^3$He(d,p)$^4$He reaction;~the dotted curve
represents $S(E)$ for the bare nuclear interaction, extracted from
the solid curve by removing the effects of the electron
shielding.}
\end{figure}

\begin{figure}
\includegraphics{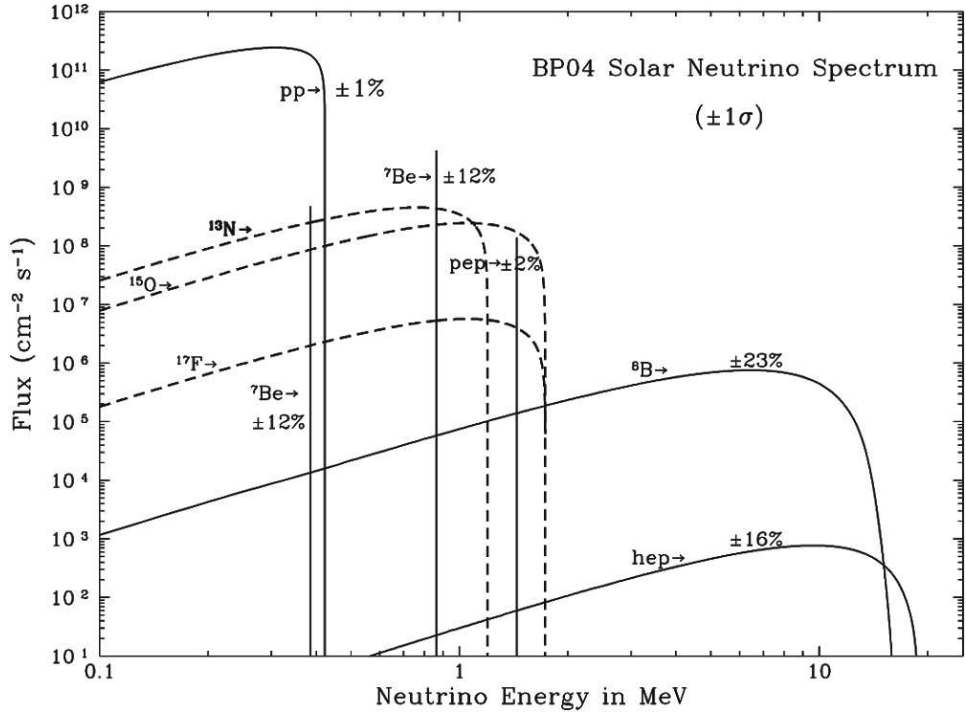}
\caption{The solar neutrino spectrum of BP04 \cite{9}.}
\end{figure}

\begin{figure}
\includegraphics{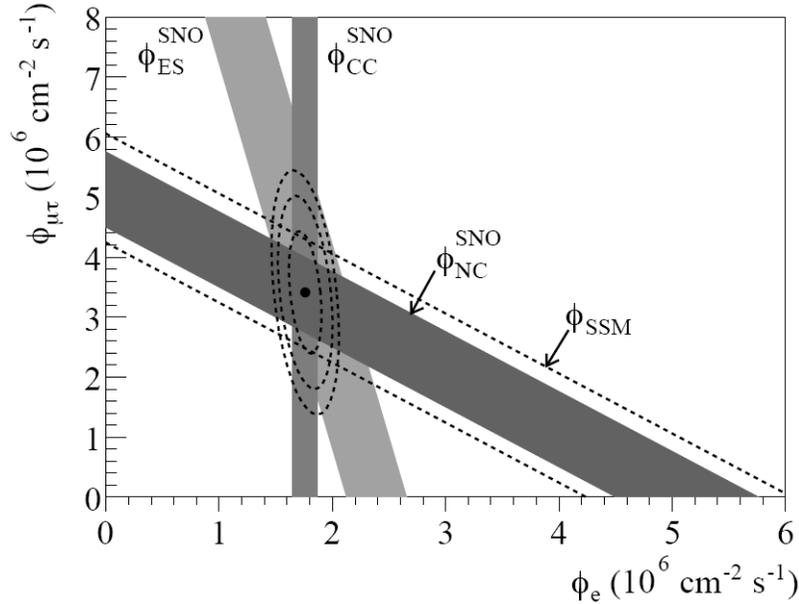}
\caption{The flux of $^8$B solar neutrinos which have either $\mu$
or $\tau$ flavor \textit{vs.}~the corresponding flux of electron
neutrinos, as deduced from the charge--current (CC),
neutral--current (NC), and elastic--scattering (ES) reactions in
the SNO heavy--water detector \cite{78}.}
\end{figure}

\end{document}